# Raman Scattering from High Frequency Phonons in Supported n-Graphene Layer Films

*A. Gupta [a], Gugang Chen[a], P. Joshi [b], S. Tadigadapa [b] and P.C. Eklund [a,c]*

[a] Department of Physics, [b] Electrical Engineering

[c] Materials Research Institute and Department of Materials Science and Engineering

The Pennsylvania State University, University Park, PA 16802 USA

**ABSTRACT**

Results of room temperature Raman scattering studies of ultrathin graphitic films supported on Si (111)/$SiO_2$ substrates are reported. The results are significantly different from those known for graphite. Spectra were collected using 514 nm radiation on films containing from n=1 to 20 graphene layers, as determined by atomic force microscopy. Both the 1st and 2nd order Raman spectra show unique signatures of the number of layers in the film. The $n$GL film analog of the Raman G-band in graphite exhibits a Lorentzian lineshape whose center frequency shifts linearly relative to graphite as ~*1/n* (for n=1 $\omega_G$~1588 cm$^{-1}$). Three weak bands, identified with disorder-induced 1st order scattering, are observed at ~ 1350, 1450 and 1500 cm$^{-1}$. The 1500 cm$^{-1}$ band is weak but relatively sharp and exhibits an interesting n-dependence. In general, the intensity of these D-bands decreases dramatically with increasing n. Three 2nd order bands are also observed (~2450, ~2700 and 3248 cm$^{-1}$). They are analogs to those observed in graphite. However, the ~2700 cm$^{-1}$ band exhibits an interesting and dramatic change of shape with n. Interestingly, for n<5 this 2nd order band is more intense than the G-band.

Several interesting reports of unusual electrical transport in ultra-thin graphite films have appeared recently [1-15]. These films are comprised of only a few atomic layers of sp2-bonded carbon known as "graphene" layers. Below, we refer to these unique atomically thin films as n-graphene layer films (i.e., $n$GL). Recently, bottom gate modulation of the current in the plane of the $n$GL film was also demonstrated which represents the first time that gate control of the current has been observed in a metallic or semi-metallic system [10]. Interesting oscillatory magneto-transport phenomena sensitive to the shape of the Fermi surface [7, 8, 10, 12, 14-17] and an unusual half-integer Quantum Hall Effect [5] have been reported [7, 9, 16-18].

Crystalline graphite is a strongly anisotropic layered solid consisting of a periodic ..ABAB.. stacking of graphene layers. All the formal chemical bonding in graphite is associated with the strong $sp^2$ intralayer bonds; these layers are, in turn, coupled by very weak van der Waals bonds. As a result of this strongly anisotropic bonding, weak (strong) electron and phonon dispersion is calculated [19, 20 21-31] and observed [32, 33] along the directions perpendicular (parallel) to the graphene planes. Any unique phonon character of *supported n*GL films is therefore expected to be due to the importance of the reduction and/or eventual loss of these interlayer forces (i.e., at n=1) and to a combination of weak charge transfer and/or strain associated with the coupling of the film to the substrate.

Phonon and electronic state dispersion in a single graphene layer has been studied theoretically for many years [34-36]. In fact, these dispersion curves led to the prediction of chirality-dependent one-dimensional electronic and phonon dispersion in carbon nanotubes [20]. Experimental success in realizing well-ordered graphene (or $n$GLs) has been somewhat limited until very recently. A few years ago, a synthesis breakthrough was reported which is generally applicable to many layered solids that exhibit weak interlayer bonds (e.g., BN, NbSe$_2$) [37]. The approach relies on the careful mechanical transfer of a thin section of material from the outer basal plane surface of a parent crystal to a supporting substrate. In one case, this was elegantly accomplished by mounting a small graphite crystal on a tip-less cantilever held in an atomic force microscope (AFM) [10]. Thin n~3 $n$GLs have also been fabricated by the preferential thermal evaporation of Si from the (0001) surface of SiC in ultrahigh vacuum and they have been shown to also exhibit oscillatory magnetoresistance [14].

Here we present the first formal report of our Raman scattering studies on $n$GLs, where n is in the range 1<n<20, as determined by AFM [38]. The films were made by mechanically transferring thin flakes of material from highly oriented pyrolytic graphite (HOPG) on to a reasonably smooth Si substrate. Despite the short range and weak interlayer forces within these films, we have observed a surprisingly continuous and systematic change in the 1$^{st}$ and 2$^{nd}$ order high frequency intralayer phonons as the number of layers $n$ in the film increases. Raman scattering can therefore be used to identify the number of layers in an $n$GL film.

Our films were prepared as follows. Thin basal plane sections were removed from large slabs (~5x5x1mm$^3$) of highly oriented pyrolytic graphite (HOPG) [39, 40] by Scotch Tape (3M) and transferred to Si (111) substrates. These substrates have a ~ 100 nm thick thermally grown oxide overlayer, such as typically used for bottom gate control of nanotubes and nanowire devices [41, 42]. An atomic force microscope (AFM; model XE 100, PSIA, Inc.) operating under ambient conditions was used to measure the surface roughness of the substrate (~1-2 nm) and the apparent thickness of the various $n$GL films studied here. Typical minimum lateral dimensions of these $n$GLs usually exceeded ~2-4 μm. MicroRaman spectra could therefore be collected from a single $n$GL using a 100x objective with a spot size of ~ 1 μ. A variety of $n$GL flakes were deposited on a single Si substrate that was in turn mounted on the precision x-y translation stage of our Raman microscope. Raman spectra were then collected using 514.5 nm excitation under ambient conditions at low laser power (<2 mW) using a JY-ISA T64000 triple holographic grating spectrometer with an L-N$_2$ cooled CCD detector. A low-pressure Hg lamp (Oriel) was set up in the background to provide a reference frequency in all spectra that is present as a narrow line at 1122.5 cm$^{-1}$. The intrinsic resolution of the Raman spectrometer was also observed and recorded as the lineshape of this Hg line. Specific $n$GLs that had been previously characterized by AFM (film thickness and basal plane shape) could be recognized in the video display of the Raman microscope by their characteristic irregular shape. In this way, we were able to collect Raman spectra of $n$GL films of known thickness ($n$).

In Fig. 1, we plot the measured AFM thickness (h) vs. our assigned value for n. The straight line is the result of a linear least squares fit to the data. In making the tentative assignment for n, we assumed that the film would have a thickness given by the linear relation h ~n*t + t$_0$, where n is an integer (number of layers), t is the approximate theoretical thickness of a graphene layer, and t$_0$ is an instrumental offset (i.e., independent of n). The data for $h$ in Fig. 1 represents the average step height obtained for a single $n$GL flake from averaging ~100 individual line scans using a Si AFM cantilever.

As can be seen, the linear fit to the *h(n)* data is very good and we obtain  t=0.35 ± 0.01 nm, in good agreement with c/2=0.335 nm for graphite [19, 20, 43]; the offset obtained is $t_0$=0.33 ± 0.05 nm.  We propose that $t_0$ may be the result, in part,  of a different force of attraction between the tip and graphene, as compared to the tip with $SiO_2$. From the least squares fit, we find that the n=1 nGL has an apparent thickness of  *h(1)=(t+t₀)=0.68 ± 0.07 nm.*

In Fig. 2, we show a series of  high frequency Raman spectra for *n*GLs of various *n* collected  in  a  single  experimental  session.   By  this  we  mean  that  the  spectra  were collected in succession by only moving the x-y stage to position a new *n*GL flake under the microscope objective.  The spectra in Fig. 2 cover the region ~1500-3600 cm⁻¹, and therefore contain contributions from both 1st order scattering, i.e., the G-band (~1582 cm⁻¹ in graphite) and also 2nd order scattering. The Raman spectra have been stacked in the figure  from  bottom  to  top  according  to  increasing  n;  the  spectrum  of  HOPG  (n~8) appears  on  top.   Furthermore,  the  intensity  of  each  spectrum  in  Fig.  2  has  been  scaled  to artificially  produce  the  same  strength  2nd  order  band  near ~ 2700  cm⁻¹.  If  plotted  in absolute Raman scattering intensity units (e.g., counts/watt-sec), the true G-band intensity in Fig. 2 would increase  over the range n=1 to ~20 almost linearly with the number of layers  in  the  *n*GL.    In  most  of  our  experiments  (passes),  the  G-band  for  HOPG  is comparable in intensity to that of n~19, suggesting that the optical penetration depth  in the *n*GL is ~ 20 layers.  Several general observations can easily be made from Fig. 2: (1) the G-band frequency downshifts with increasing n (for small n), (2) the 2nd order ~2700 cm⁻¹ band exhibits an interesting *n*-dependence in shape and  frequency, (3) the sharp 2nd order feature at ~3248 cm⁻¹ is insensitive to *n*, (4) for n<5, the 2nd order ~2700 cm⁻¹ band for an $n$GL is more intense than the 1$^{st}$ order G-band, and (5) additional weak features are present in the first order region ($\omega < 1620$ cm$^{-1}$) (we show these later on an expanded intensity scale).

In Fig. 3, on an expanded frequency scale, we show the G-band for various nGLs. The spectra are stacked from bottom to top with increasing n. The spectra in the figure were also collected in a single pass, i.e., they were also collected sequentially under identical optical conditions. The Raman spectra in Fig. 3 are not scaled, although the intensity of the HOPG G-band is less than normally observed (the HOPG G-band intensity is typically comparable to n~19 data). As can be seen in the figure, the G-band clearly downshifts with increasing n. Careful Voight lineshape analyses of these G-bands show that they are extremely well fit by a single Lorentzian that has been "folded" with a Gaussian instrument function. The Gaussian is defined by the spectral shape of the low-pressure Hg line present in all our Raman spectra. Interestingly, the G-band frequency $w_G$ exhibits an almost linear dependence on 1/n, i.e., $w_G(n) = w_G(8) + b/n$, where $\beta \sim 6$ cm$^{-1}$ is a constant. This behavior is shown in Fig. 4 which presents the results of 8 separate studies (passes) mostly, but not exclusively, obtained from the same set of $n$GLs. However, in each successive pass, a different location on the particular nGL was excited by the laser.

It is of interest to compare the value obtained for the G-band frequency of SiO$_2$-supported graphene, i.e., $w_G(1)=1587$ cm$^{-1}$ (Fig. 4) to the strongest high frequency 1$^{st}$ order mode of moderate diameter (e.g.,D~1.4 nm) single-walled carbon nanotubes (SWNT), e.g., $\omega \sim 1592$ cm$^{-1}$ [19, 20, 44]. For SWNTs near this diameter (e.g., 1.2 -1.6 nm), it is thought that the wall curvature of the nanotube does not strongly effect the C-C intra-atomic force constants [20, 45]. If we accept this point of view, then our value for $\omega_G(n=1) \sim 1587$ cm$^{-1}$ is almost 5 cm$^{-1}$ less than anticipated.

Our Voight lineshape analysis of the G-bands typically returned an uncertainty for the band frequency of only $\sim 0.1$-$0.2$ cm$^{-1}$. Furthermore, absolute frequency calibration uncertainty in our experiments is small ($\sim 0.1$-$0.2$ cm$^{-1}$), as all G-band frequency scales are referenced to the nearby Hg line. Thus, it is somewhat surprising that the scatter in the frequency ($\omega_G$) data (Fig. 4) for fixed n is about twice as large as anticipated, i.e., the observed scatter is $\sim \pm 0.5$-$1$ cm$^{-1}$. The larger scatter in $\omega_G(n)$ data suggests that there may be a spot-to-spot variation within the nGL films. Perhaps this is due to the spectral difference produced by excitation near (or far) from a grain-boundary in the HOPG (the HOPG basal plane grain size is $\sim 1$-$3$ $\mu$ which is comparable to the laser spot diameter).

The symmetry and degeneracy of the long wavelength (q=0) optical phonons in graphite and graphene can be obtained by a group theoretical analysis. For graphite, with four atoms per primitive cell (two per layer in adjacent A and B layers), it has been shown that the atomic representation reduces to $\chi_{graphite} = E_{2g}^{(1)} + E_{2g}^{(2)} + E_{1u} + A_{1u} + B_{1g}$, where the E modes are doubly degenerate, and the A and B modes are singly degenerate. The $E_{2g}$ symmetry modes are Raman-active ( $\omega(E_{2g}^{(1)}) = 42$ cm$^{-1}$; $\omega(E_{2g}^{(2)}) = 1582$ cm$^{-1}$). The low frequency $E_{2g}^{(1)}$ modes can be identified with a rigid shearing of adjacent graphene layers [20]. The low frequency of this mode is a measure of the weak interlayer van der Waals forces. On the other hand, the high frequency Raman-active $E_{2g}^{(2)}$ modes and IR-active $E_{1u}$ (1587 cm$^{-1}$) modes both involve a stretching of unusually strong *intra*layer bonds; therefore only a small component of the mode restoring force comes from *inter*layer bonds. The out-of-plane graphite $A_{1u}$ (868 cm$^{-1}$) mode is IR-active and the $B_{1g}$ mode (127 cm$^{-1}$) is optically silent. Both the electronic states and the phonon states must depend, to some extent, on $n$ in an $n$GL. Of course, the question remains: "How much do these states vary with n for small n, and with increasing n, by what value of n will the variation cease to be important"?.

For the case of graphene, there are 2 atoms per primitive cell, and it can be easily shown that $\chi_{graphene} = E_g + A_1 + B_1$. In graphene, it can then be seen that several "graphitic" modes are lost, i.e., their re-appearance in an $n$GL requires at least a doubling of the number of layers (n=2) and an interlayer interaction. Several calculations of the phonon modes for graphene have appeared [34-36, 46]. Unfortunately, to date, there have not been any experimental reports to test the subtle differences that can occur between the phonons in graphene and graphite. The phonon dispersion and corresponding one-phonon DOS for graphene calculated according to Mapeli and collaborators [ref] appears in Fig. 5. Particular attention in the figure should be paid to strong peaks in the DOS at ~ 1380 cm$^{-1}$ and 1600 cm$^{-1}$ and the weak peak at ~ 1500 cm$^{-1}$. Evidence for structure in the 1st and 2$^{nd}$ order Raman spectra associated with these maxima are seen, as discussed below

From the symmetry analysis, it should be clear that the Raman band we observe in the interval ~1581-1588 cm$^{-1}$ in $n$GLs (Fig'2 and 3) should be connected with the $E_{2g}^{(2)}$ modes of graphite (G-band), as perturbed by the small number of layers in the film and the coupling of the film to the substrate. However, the one-phonon scattering symmetry analysis does not allow for the weak Raman bands that we observe in $n$GLs near ~ 1350, 1450 and 1500 cm$^{-1}$. These weak Raman bands can be seen in Fig. 6, where we have considerably expanded the intensity so that the G-band maxima are way off-scale. The strongest band in the figure (~1350 cm$^{-1}$) is identified with a "disorder"-induced band or "D-band", as it is known in the literature, that has been observed in many sp2-bonded carbons. These studies have shown that the Raman D-band intensity and position, respectively, depend on the degree and nature of the basal plane disorder and the excitation wavelength [20, 31, 47-52]. For the n=1 *n*GL (Fig. 6), we observe that the D-band intensity $I_D$ of the ~1350 cm$^{-1}$ band is ~ 1/5 of that of the G-band ($I_G$). However, with increasing n, the D-band intensity $I_D$ can be seen to dramatically decline relative to $I_G$. We observe that the intensity ratio ($I_D/I_G$) exhibits an approximately exponential decline with increasing n In addition to the ~1350 cm$^{-1}$ D-band, two additional weak bands are observed that are also assigned to disorder-induced 1$^{st}$ order scattering or "D-scattering": ($D_2$) ~1450 cm$^{-1}$ and ($D_3$)1500 cm$^{-1}$. It should be noted that the graphene phonon calculations of Mapeli et al. predict a weak but sharp DOS peak at ~1500 cm$^{-1}$. This 1500 cm$^{-1}$ peak is particularly interesting in that it has striking n-dependence, whereas the other D-scattering peaks (e.g., $D_2$~1350 cm$^{-1}$ and $D_3$~1450 cm$^{-1}$) are largely insensitive to n.

It is also interesting to speculate about the origin of the *n*-dependence of the D-band scattering in nGLs. One explanation that we might offer for this observation is a bending of the graphene layers necessary to accommodate the surface roughness of the substrate. This view is motivated by the fact that our SiO2:Si substrates are not atomically smooth. Our AFM measurements indicate a substrate roughness of ~ 1-2 nm, i.e., ~3-6 graphene layers. The n-dependence of the D-scattering then may stem from the loss in sp2 bond bending disorder associated with the increase in rigidity of the nGL as the number of layers n increases. This cooperative (multilayer) film rigidity is expected to resist the deformation from planarity as the film, via a van der Waals coupling between substrate and $n$GL, attempts to conform to the underlying substrate roughness. That is, an n=1 nGL is relatively very compliant and bends easily, while an n=5 nGL should be much more rigid. Our suggestion of a planar deformation to induce D-band scattering, although novel, is consistent with the general concept that many different mechanisms that break the translational symmetry of the basal plane can give rise to D-band scattering (e.g., finite basal plane size, boron-doping, bond-disorder, etc.) [31, 47-53]. Further work will be needed to confirm that variable film rigidity is responsible for the D-band scattering. This idea could be experimentally tested by studying films supported on atomically smooth substrates (e.g., mica) or suspended over a small hole in a substrate.

In Fig. 7, we display in three separate panels, the 2nd order features shown in Fig. 2, but on an expanded intensity scales. For clarity, the spectra are normalized to give them all approximately the same intensity, independent of n. The panels each display the evolution (n=1 (bottom) to HOPG (top)) of one of the three 2nd order bands observed at: ~2450 cm$^{-1}$ (weak intensity), ~2700 cm$^{-1}$ (moderate to strong intensity) and ~3248 cm$^{-1}$ (moderate intensity). Based on the peaks in the phonon DOS of Mapeli et al. (Fig. 5), the ~2450 cm$^{-1}$ and ~2700 cm$^{-1}$ bands must due to combination scattering, i.e., ($\omega_1+\omega_2$), while the relatively narrow 2nd order band at ~3248 cm$^{-1}$ must be a 2$\omega$ feature (overtone scattering) associated with a doubling of the highest phonon frequency in the zone. It is interesting that although we observe that the G-band (q=0 mode of this optical branch) upshifts with decreasing n, while the maximum frequency of the same branch $\omega$(max) remains fixed in frequency. This result, amongst other experimental observations presented here, seems to be a good test for future lattice dynamics models of $n$GLs and *supported $n$*GLs..

In summary, we have prepared a variety of ultrathin graphitic films (or $n$Gls) by mechanical transfer from an HOPG slab to a $SiO_2$:Si substrate. The film thickness of various $n$GL flakes was measured by averaging AFM z-scans. They exhibited lateral dimensions on the order of ~2-5 microns The effective film thickness data $h$ vs. *assigned n* was found to be well fit to a straight line, yielding the value t=0.355 for the thickness of a graphene layer, a little larger than c/2 for graphite, but within experimental error. MicroRaman spectra were then collected on these films and correlated with the integer number of graphene layers per film. Surprisingly, despite the anticipated short range of the c-axis forces in graphite, the Raman spectra of nGLs were found to be very sensitive to n, for n as large as n=10. Strong signatures of the number of graphene layers per film are seen in the weak $D_3$-scattering peak at ~1500 $cm^{-1}$, in the position of the G-band, and in the $2^{nd}$ order band observed near ~2700 $cm^{-1}$. Detailed lattice dynamics calculations should be carried out on atomically thin films (vs. n) to understand how the subtle changes we observe in zone-center, mid-zone and zone edge modes affect the D, $1^{st}$ and $2^{nd}$ order scattering. It will also be necessary to extend these experimental studies to other excitation wavelengths ($\lambda_{ex}$) to see which of these Raman bands are dispersive (i.e., the position of dispersive Raman bands depends on $\lambda_{ex}$). The simple 1/n behavior of the G-band position also remains unexplained. It is not known to what extent the 1/n dependence is connected to strain (via coupling to the substrate) or to the finite number n of C-layers in the film.

Finally, the $n$GL D-band scattering intensity (~1350 $cm^{-1}$) is observed to strongly decrease with increasing n. We have proposed that this behavior might correlate with the rigidity of the film and its ability to avoid distortion to accommodate the substrate roughness.

**Acknowledgements:** This work was supported, in part, by Penn State Materials Research Institute and the Penn State MRSEC under NSF grant DMR 0213623. Authors acknowledge useful discussion with Dr. Humberto R. Gutierrez (Penn State) and Cynthia Coggins (PSIA, Inc.)

**Figure captions**

Fig. 1.   Effective $n$GL film height vs. assigned n.  The straight line is a least square fit to the data.  The apparent thickness of a graphene layer is t=0.35±0.01 nm, and the AFM offset parameter is $t_0$=0.33 nm (see text for discussion).

Fig. 2.   High-frequency 1st and 2nd order microRaman spectra of $n$GL films supported on a SiO2:Si substrate and HOPG. The data were collected using 514.5 nm radiation under ambient conditions.  The spectra are scaled to produce an approximate match in intensity for the ~2700 cm$^{-1}$ band.  Note: (1) the shape and frequency of this 2nd order band is sensitive to the number of layers n and (2) the relative intensity of the 2nd order band at 2700 cm$^{-1}$ is larger than the 1st order-allowed G-band for n<5.

Fig. 3.   1st order-allowed Raman G-band for supported nGL films vs. number of layers n. The data were collected using 514.5 nm radiation under ambient conditions.  The position of this band upshifts linearly relative to graphite with increasing 1/n (see Fig. 4).

Fig. 4.   G-band frequency  vs. 1/n.  Each pass represents a set of data collected under identical optical conditions.  The solid line represents the results of a least squares fit to the data: $\omega_G$(infinity)=1581.6 cm$^{-1}$ and d$\omega$/d(1/n)=5.5 cm$^{-1}$.  The average of five values found for supported graphene (n=1) is $\omega_G$ =1587.1 cm$^{-1}$.

Fig. 5.   Calculated phonon dispersion curves and one-phonon density of states (DOS) for unsupported (free standing) graphene by Mapeli et al. [46]. Structure in the phonon DOS is expected in D- and 2nd order-scattering.

Fig. 6.   Raman spectra (expanded intensity scale) of supported $n$GL films and HOPG in the vicinity of the G-band. The data were collected using 514.5 nm radiation under ambient conditions. The spectra have been scaled to show the weak structure which has been identified with the one-phonon DOS in Fig. 5. The arrows in the figure identify weak scattering features.  Note that the $D_3$ feature at ~ 1500 cm$^{-1}$ is sensitive to the number of layers n.

Fig. 7  2nd order Raman spectra of supported nGL films.  The data were collected using 514.5 nm radiation under ambient conditions. The spectra in each panel have been scaled to generate features of approximately the same height. These 2nd order bands also appear in Fig. 2 which indicates their approximate relative scattering intensity to the G-band. Note that the ~ 2700 cm$^{-1}$ band is sensitive to the number of layers n.

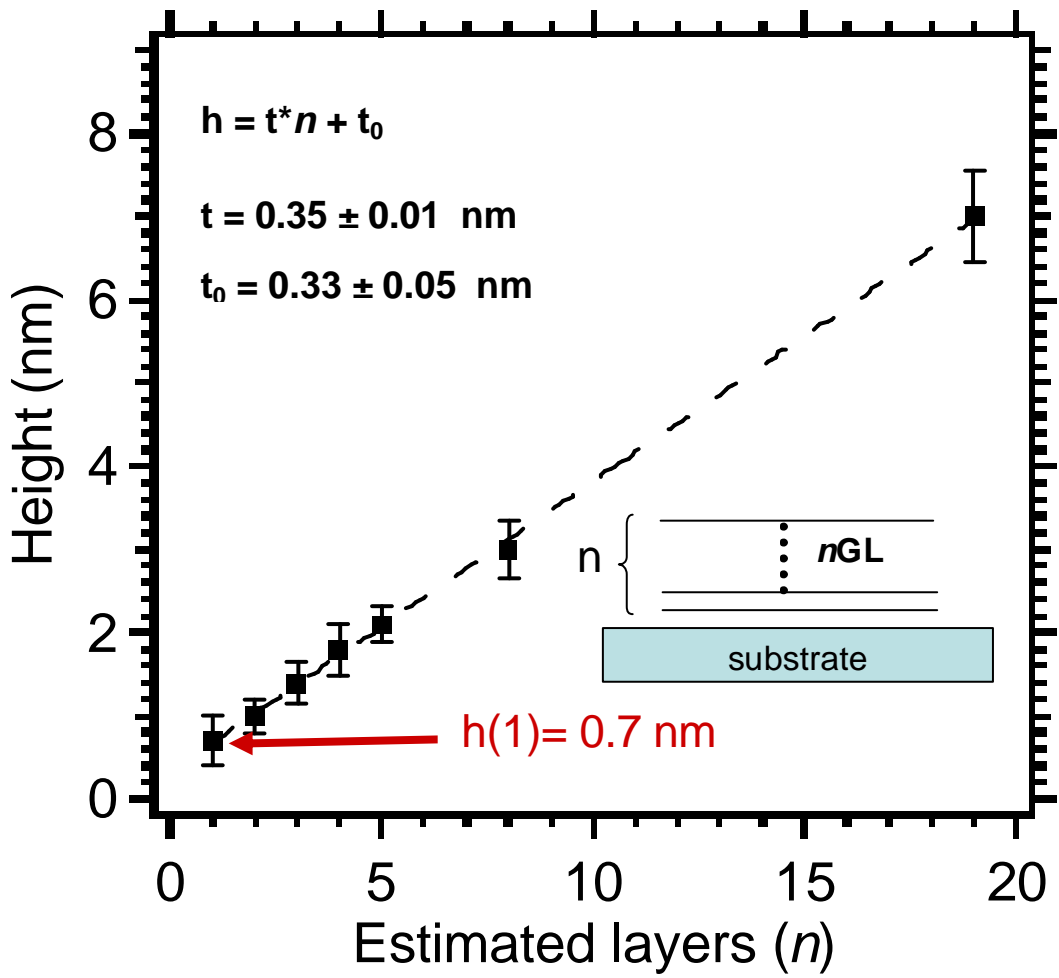

**Fig. 1**

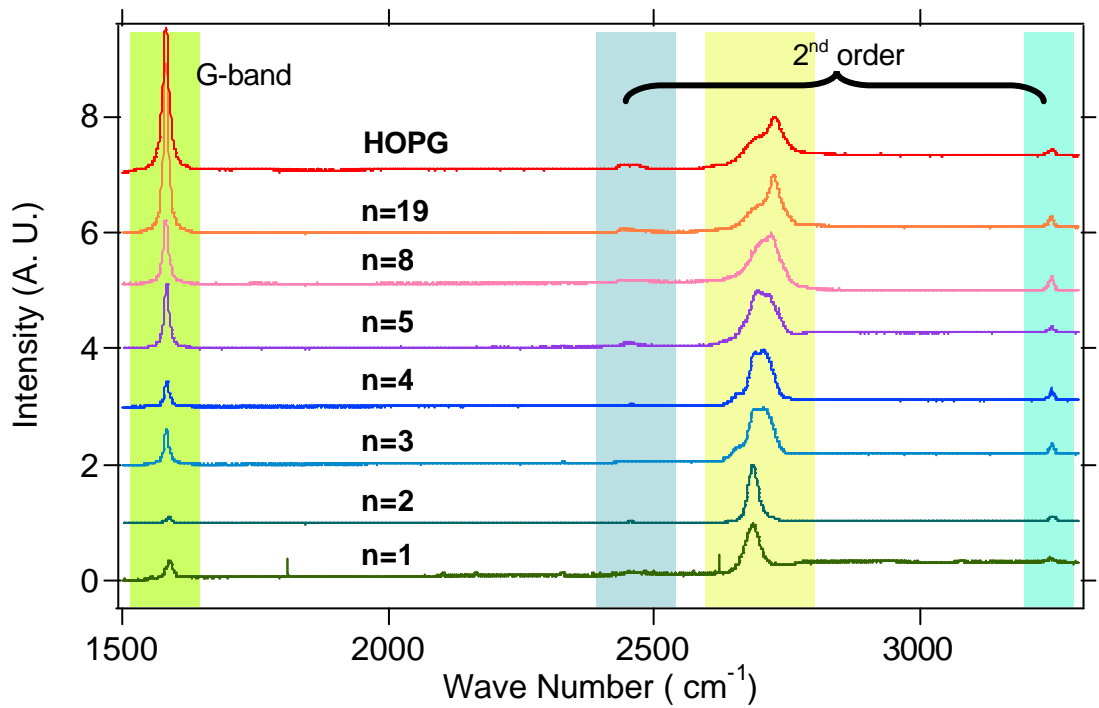

**Fig. 2**

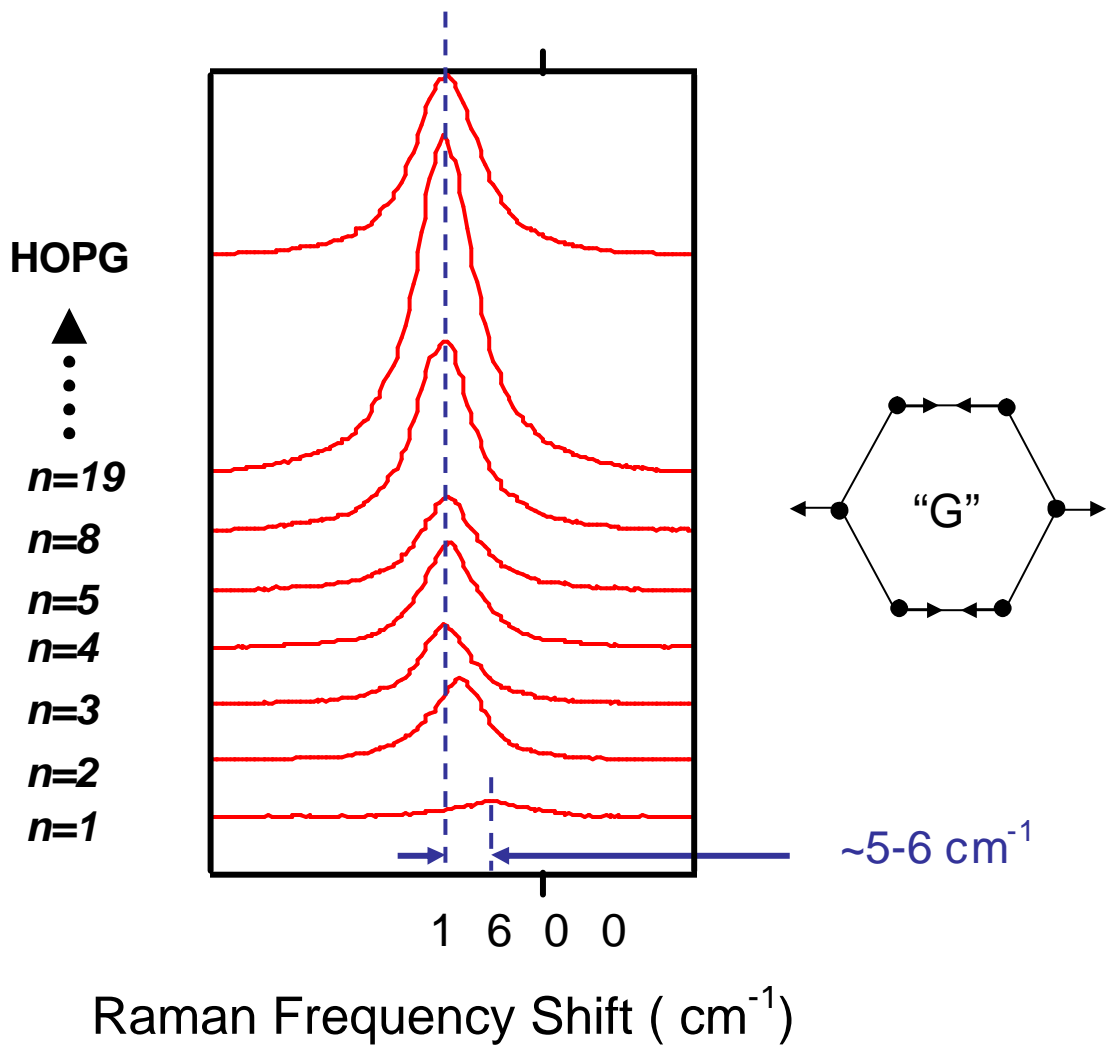

**HOPG**

▲
•
•
•

*n=19*

*n=8*

*n=5*

*n=4*

*n=3*

*n=2*

*n=1*

~5-6 cm$^{-1}$

"G"

1 6 0 0

Raman Frequency Shift ( cm$^{-1}$)

**Fig. 3**

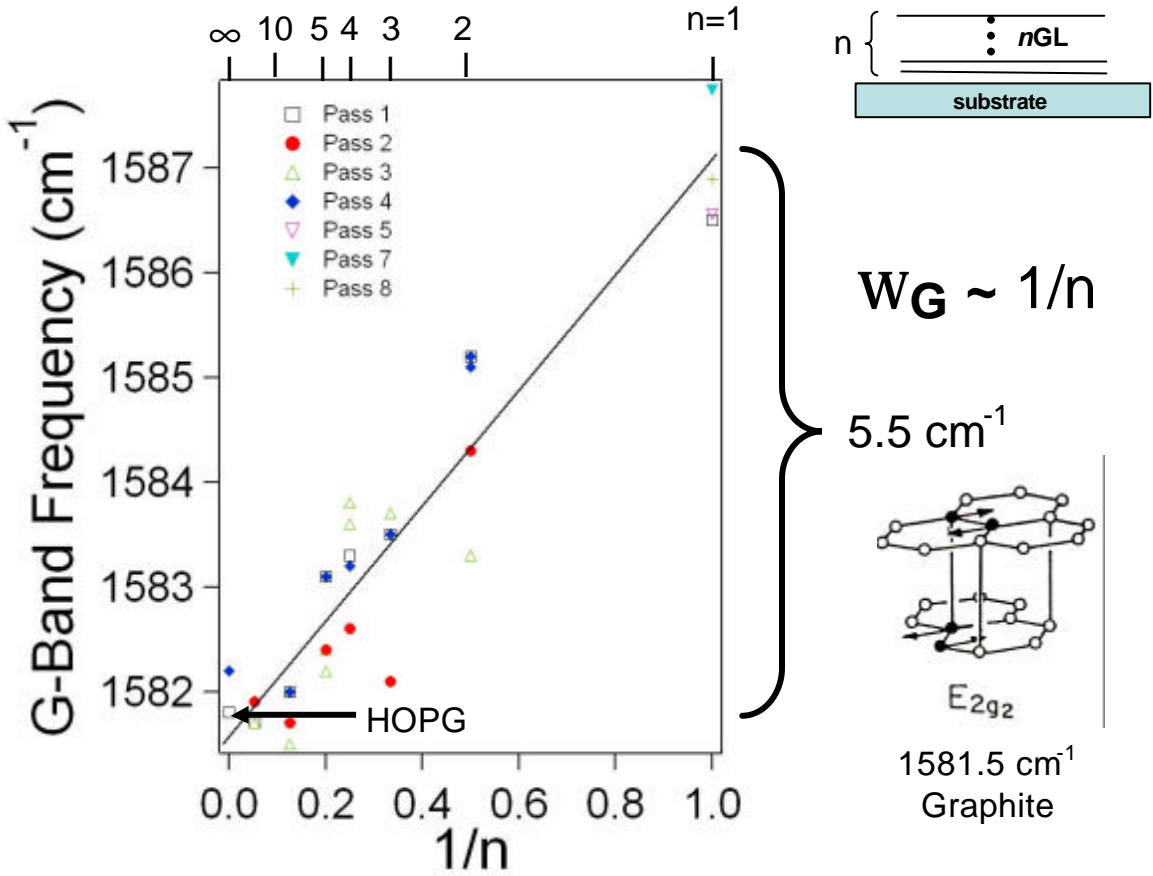

**Fig. 4**

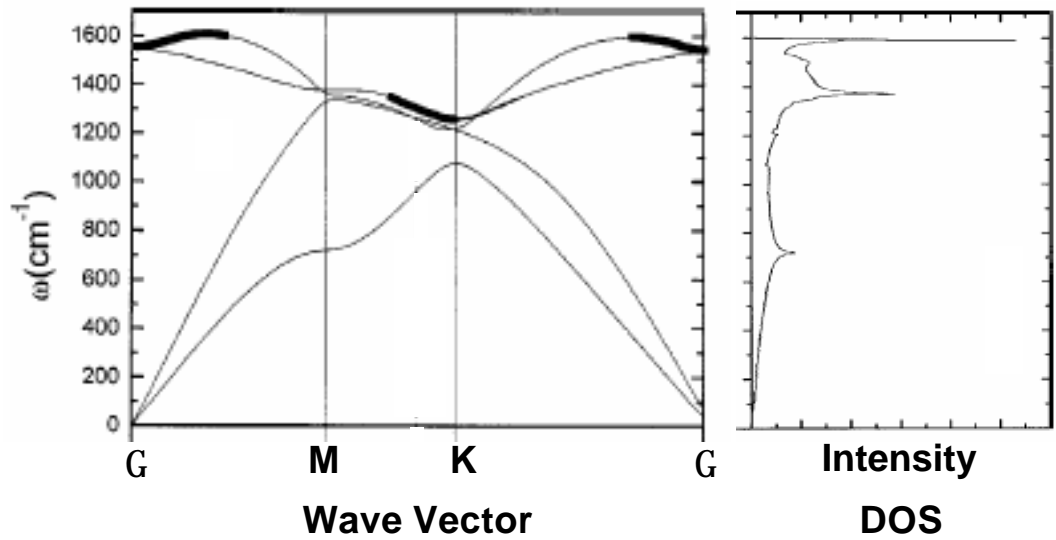

**Fig. 5**

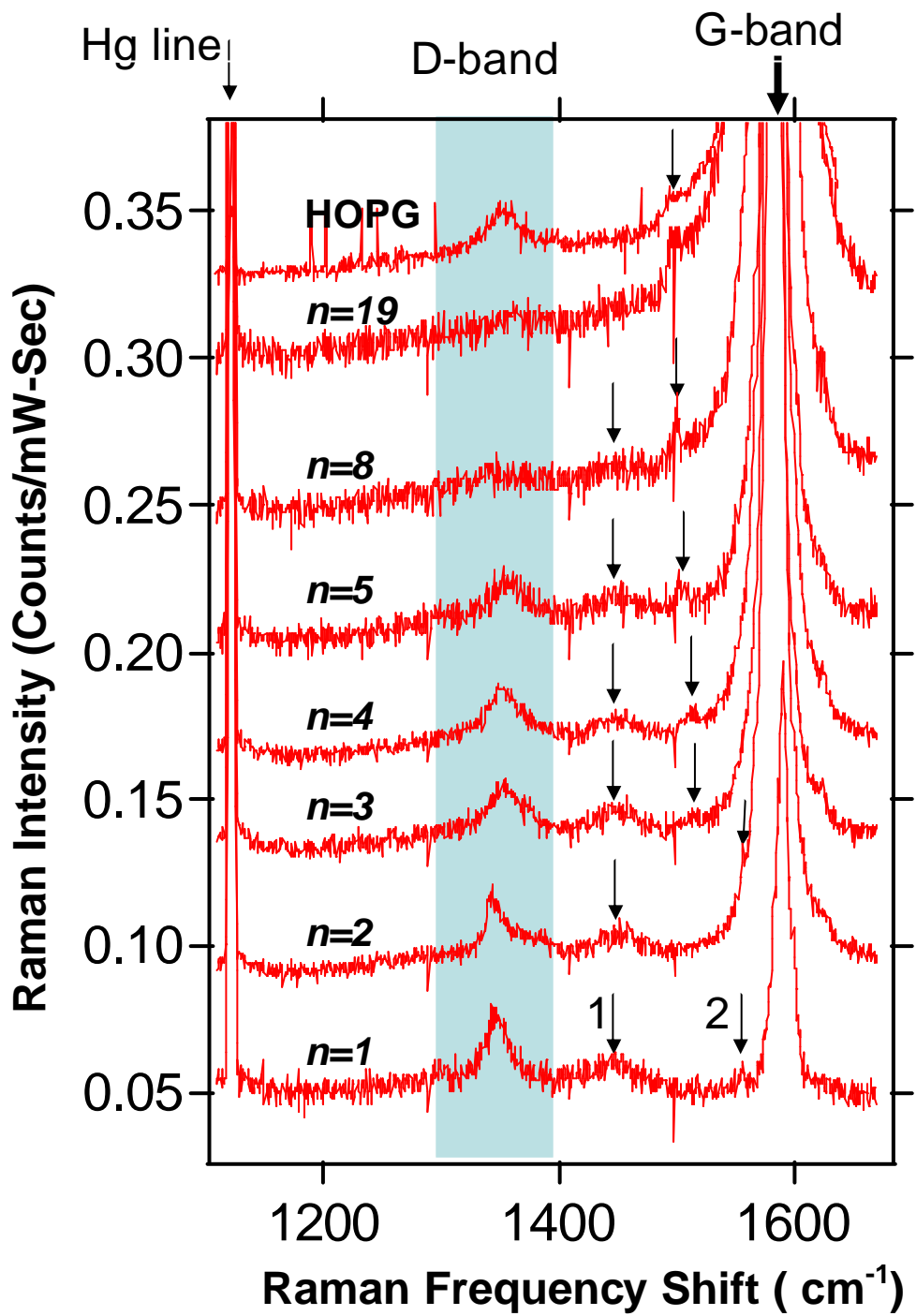

Fig. 6

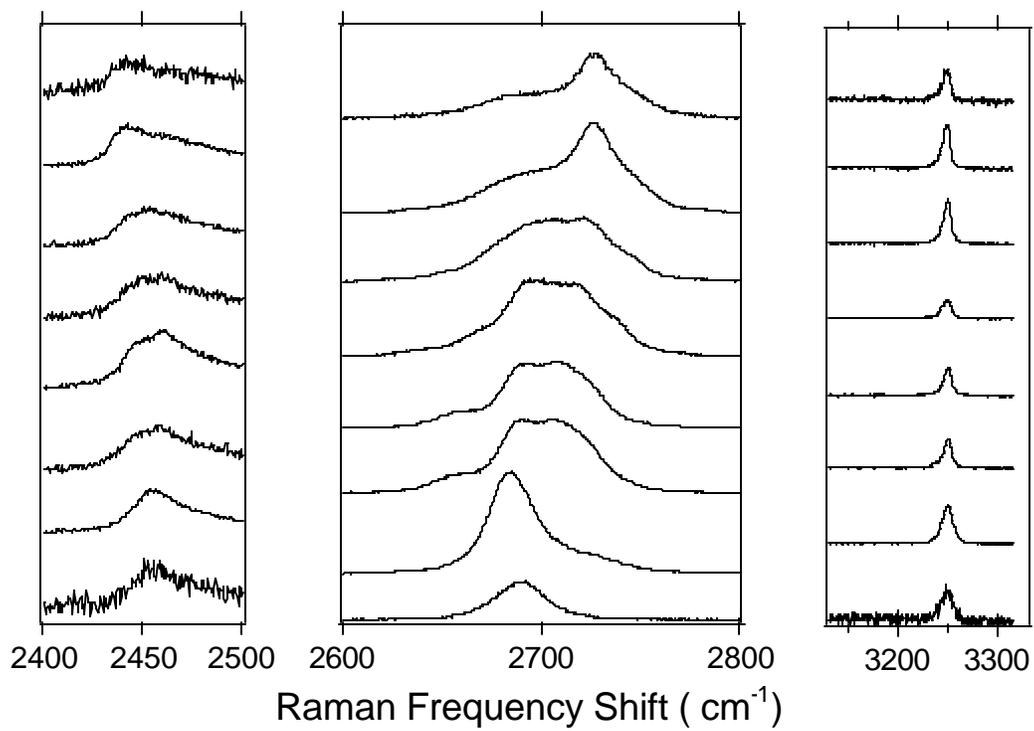

Raman Frequency Shift ( cm⁻¹)

**Fig. 7**